\documentclass[aps,twocolumn,superscriptaddress]{revtex4}
\usepackage{graphicx}
\usepackage{qcircuit}

\newcommand{\ket}[1] {\left| #1 \right\rangle}

\begin{document}

\title{Numerical analysis of quantum circuits for state preparation and unitary operator synthesis}

\author{Sahel Ashhab}
\affiliation{Advanced ICT Research Institute, National Institute of Information and Communications Technology (NICT), 4-2-1, Nukui-Kitamachi, Koganei, Tokyo 184-8795, Japan}

\author{Naoki Yamamoto}
\affiliation{Quantum Computing Center, Keio University, 3-14-1 Hiyoshi, Kohoku-ku, Yokohama, Kanagawa 223-8522, Japan}
\affiliation{Department of Applied Physics and Physico-Informatics, Keio University, Hiyoshi 3-14-1, Kohoku-ku, Yokohama 223-8522, Japan}

\author{Fumiki Yoshihara}
\affiliation{Advanced ICT Research Institute, National Institute of Information and Communications Technology (NICT), 4-2-1, Nukui-Kitamachi, Koganei, Tokyo 184-8795, Japan}
\affiliation{Department of Physics, Tokyo University of Science, 1-3 Kagurazaka, Shinjuku-ku, Tokyo 162-8601, Japan}

\author{Kouichi Semba}
\affiliation{Advanced ICT Research Institute, National Institute of Information and Communications Technology (NICT), 4-2-1, Nukui-Kitamachi, Koganei, Tokyo 184-8795, Japan}
\affiliation{Institute for Photon Science and Technology, The University of Tokyo, 7-3-1 Hongo, Bunkyo-ku, Tokyo 113-0033, Japan}

\date{\today}

\begin{abstract}
We perform optimal-control-theory calculations to determine the minimum number of two-qubit CNOT gates needed to perform quantum state preparation and unitary operator synthesis for few-qubit systems. By considering all possible gate configurations, we determine the maximum achievable fidelity as a function of quantum circuit size. This information allows us to identify the minimum circuit size needed for a specific target operation and enumerate the different gate configurations that allow a perfect implementation of the operation. We find that there are a large number of configurations that all produce the desired result, even at the minimum number of gates. We also show that the number of entangling gates can be reduced if we use multi-qubit entangling gates instead of two-qubit CNOT gates, as one might expect based on parameter counting calculations. In addition to treating the general case of arbitrary target states or unitary operators, we apply the numerical approach to the special case of synthesizing the multi-qubit Toffoli gate. This approach can be used to investigate any other specific few-qubit task and provides insight into the tightness of different bounds in the literature.
\end{abstract}

\maketitle

\section{Introduction}
\label{Sec:Introduction}

Quantum computing devices are making rapid progress towards large-scale practical applications \cite{Ladd,Buluta}. Presently available devices have been used to demonstrate quantum advantage, in which a quantum computer performs a computational task faster than the fastest present-day classical computer \cite{Arute,Wu}.

The standard approach to performing quantum protocols is the so-called circuit model. In this approach a sequence of quantum gates is applied to the initial quantum state, concluding with a measurement of the final state to extract the output of the computation. The sequence of quantum gates used to perform the algorithm or any part of it is sometimes called the quantum circuit.

It is well known that any unitary operator can be decomposed into, or in other words synthesized from, a sequence of single- and two-qubit gates, provided that the elementary gate set is universal \cite{Barenco,Knill,ShendeUnitary,Bergholm,Mottonen,Nielsen}. Similarly, any desired quantum state can be prepared from any initial state using a sequence of elementary gates \cite{Grover,Kraus,Kaye,ShendeStatePrep,Soklakov}. As a result, any quantum algorithm can be implemented by performing a sequence of these elementary gates applied to a standard initial state. While the natural gate set depends on the specific technology used in a given realization of the qubits and their coupling mechanism, the most common elementary gate set used in the quantum information theory literature is the CNOT gate and the set of all single-qubit unitary operators. We shall use this elementary gate set as the standard one for most of our calculations in this work.

Quantum algorithms are often designed with a block operation, e.g.~a black-box operation or oracle, that transforms a multi-qubit system in a certain desired way. It is important for practical applications to be able to decompose such multi-qubit operations into single- and two qubit gates. A large amount of literature has been devoted to the question of quantum circuit complexity, i.e.~the smallest number of single- and two-qubit gates needed to perform specific tasks in quantum computation. There are two main approaches in the study of circuit complexity. In one approach, a number of studies in the literature have proposed systematic methods to construct $n$-qubit operations from elementary gates using step-by-step recipes \cite{Znidaric,Plesch,Vidal,Vatan,Vartiainen,GoubaultDeBrugiere2020,Rakyta}. In the other approach, some studies have derived lower and upper bounds for the minimum number of gates needed to perform $n$-qubit tasks. Importantly, there are cases for which there is a gap between the gate counts obtained from the theoretical lower bounds and those obtained from recipe-based constructions in the literature.

In this work we investigate this gap by numerically calculating the minimum number of gates needed to perform a given $n$-qubit task. Our numerical results therefore provide tight lower bounds for the quantum circuit sizes needed to perform various few-qubit tasks. We find in particular that it is possible to achieve the theoretical lower bounds based on parameter counting in some cases, while there are cases in which the actual minimum circuit sizes are higher than those obtained from parameter counting calculations. Similarly, we find that quantum-circuit-construction recipes in the literature are optimal in some cases but not others. We obtained these results by modifying an optimal-control-theory algorithm for optimizing control fields such that it can be used to optimize unitary operators. Considering the exponential growth of resources needed in general to implement state preparation and unitary operator synthesis \cite{Knill,ShendeUnitary}, our numerical calculations were necessarily limited to small numbers of qubits and short gate sequences. Specifically, we found that we can investigate state preparation for up to four qubits and unitary operators for up to three qubits in the most general case before the computation time makes our numerical approach unfeasible. We therefore cannot draw definitive conclusions about large systems. However, our results on small systems can give an idea about how close past results in the literature, including both the systematic constructions and the mathematical bounds, are to the minimum number of gates needed to perform quantum computing tasks in these systems. Furthermore, our results on few qubit systems can be especially useful for noisy intermediate-scale quantum (NISQ) devices available today, since these devices are often operated at the limits of their computational power and any gate optimization can increase the complexity of problems that they can handle.

\section{Background and analysis methods}
\label{Sec:Background}

A quantum circuit is composed of a sequence of quantum gates. Typically the gate sequence contains single-qubit and entangling gates. Single-qubit gates are commonly assumed to have little cost in terms of the needed resources, e.g.~implementation time \cite{Ladd,Buluta}. The entangling gate count is therefore used as the key metric. We shall follow this convention in this work, and we shall refer to the number of entangling gates in a quantum circuit as the circuit size.

\subsection{Related recent work}

In addition to the theoretical approaches mentioned in Sec.~\ref{Sec:Introduction}, a few recent studies have also used numerical optimization techniques for quantum circuit design. Goubault de Brugi\`ere {\it et al.} \cite{GoubaultDeBrugiere2019} used numerical optimal control methods to find optimal quantum circuits for synthesizing general unitary operators using single-qubit gates and the M\o lmer-S\o rensen (MS) gate. Since the MS gate, which simultaneously couples all qubits in the system and contains a continuously tunable parameter, is the only entangling gate in the gate set, the overall structure of the quantum circuit is fixed, and only the continuous parameters of the different gates need to be determined. In this case, conventional optimal control methods can be applied. A related earlier proposal for performing quantum algorithms using the MS gate and single-qubit gates was studied by Martinez {\it et al.} \cite{Martinez}. Cerezo {\it et al.} \cite{Cerezo} proposed a variational state preparation algorithm, where the parameters of a quantum circuit are optimized with the goal to approach a target state. Shirakawa {\it et al.} \cite{Shirakawa} used numerical methods to optimize quantum circuits by adding gates one at a time and optimizing each gate while keeping the rest of the quantum circuit fixed. The authors demonstrated that this approach performs well in some cases, e.g.~for preparing ground states of physical Hamiltonians, while it does not produce optimal quantum circuits in other cases.

\subsection{Lower bounds}

\begin{figure}[h]
\includegraphics[width=8.0cm]{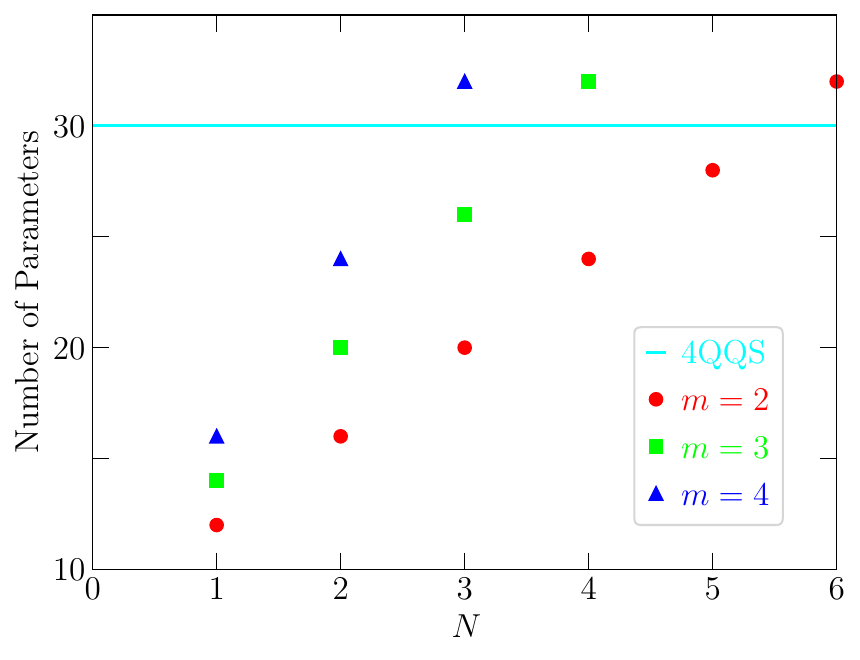}
\caption{The number of independent parameters in a four-qubit quantum state (4QQS; horizontal cyan line) and the number of independent parameters in a quantum circuit as a function of circuit size $N$. The red circles, green squares and blue triangles correspond, respectively, to using two-, three- and four-qubit CZ gates along with single-qubit rotations in constructing the quantum circuit. The number of independent quantum circuit parameters obviously increases with circuit size. The lower bound for the circuit size needed to perfectly prepare an arbitrary four-qubit state is defined by the minimum value of $N$ for which the data point lies above the horizontal line.}
\label{Fig:FidelityLowerBounds}
\end{figure}

Before describing the details of our work, we review past theoretical lower bounds for the CNOT gate counts for state preparation and unitary operator synthesis in an $n$-qubit system \cite{ShendeUnitary,Plesch}. To slightly simplify the argument, we use controlled-Z (CZ) gates instead of CNOT gates. These two gates are equivalent to each other up to single-qubit operations, and we shall refer to them by the two names interchangeably. A quantum circuit with $N$ CZ gates has $N+1$ layers of single-qubit gates, to which we also refer as rotations. At first sight, it might seem that the number of single-qubit rotations in the quantum circuit is given by the product $n\times (N+1)$. However, if two or more single-qubit rotations are applied to the same qubit in succession without this qubit being involving in entangling gates, the single-qubit rotations can be combined into one net rotation. We therefore only need to keep $n+2N$ single-qubit rotations, $n$ rotations at the initial time and 2 rotations after each CZ gate, i.e.~one rotation for each qubit involved in the CZ gate. Each single-qubit rotation is defined by three parameters that define a rotation in three dimensions. However, not all of these parameters lead to independent variations in the operation of the quantum circuit. In particular, a single-qubit rotation can be decomposed into a rotation about the z axis followed by a rotation about an axis in the xy plane. The z-axis rotation commutes with the CZ gate. As a result, z-axis rotations can be extracted from all single-qubit rotations and moved past the CZ gates to earlier steps in the quantum circuit. Apart from the first layer of single-qubit rotations, each single-qubit rotation is about an axis in the xy plane and is specified by two independent parameters. In the case of state preparation, even the first-step single-qubit rotations can be decomposed into z-axis rotations followed by xy-plane rotations, and the z-axis rotations can then be ignored because they correspond to irrelevant overall phase factors. As a result, for state preparation, the number of independent parameters is $2n+4N$. For unitary operators, the z-axis rotations in the first step of the quantum circuit cannot be ignored, which gives a total of $3n+4N$ independent parameters. The theoretical lower bound for the number of CZ gates is obtained by requiring that the number of independent parameters in the quantum circuit matches or exceeds the number of independent parameters in the target. For state preparation, the number of independent parameter defining an $n$-qubit state is $2\times 2^{n}-2$. For unitary operators the number of independent parameters is $4^n-1$. The above argument can be generalized to the case where two-qubit CZ gates are replaced by $m$-qubit CZ gates. Excluding the first step of the quantum circuit, each step contains $m$ single-qubit rotations. As a result the number of independent parameters in the quantum circuit is $2n+2mN$ for state preparation and $3n+2mN$ for unitary operator synthesis. The lower bounds are then given by $N=\lceil (2^{n}-1-n)/m \rceil$ for state preparation and $N=\lceil (4^n-1-3n)/2m \rceil$ for unitary operator synthesis, where $\lceil x \rceil$ denotes the smallest integer larger than $x$, i.e.~the ceiling function. Figure \ref{Fig:FidelityLowerBounds} shows an example of how the above parameter-counting calculations give lower bounds for the quantum circuit size needed to perfectly perform a target task.

\subsection{Target selection}

As the target state or unitary operator, we first consider random targets to investigate the most general case of an arbitrary target. To generate a random $n$-qubit state, we choose $2^n$ random complex numbers with each real or imaginary part chosen randomly from a uniform distribution centered at zero. We then calculate the norm of the resulting state vector and renormalize the state. For unitary operators, we first generate the first column of a $2^n \times 2^n$ matrix as described above for the generation of a random state. Next we similarly generate the second column, but before renormalizing the second column we subtract from it its projection on the first column to make sure that the two columns correspond to orthogonal states. We keep adding columns that are normalized and orthogonal to all previous columns until we fill the matrix. Once the matrix is complete, we randomly shuffle its columns. In spite of relying heavily on randomly generated numbers, the procedures described above for generating random quantum states and unitary operators are in fact somewhat biased in the sense that they do not produce results that are uniform in the Haar measure \cite{Stewart}. To reduce the bias, we perform an additional randomization step: we generate ten random unitary operators following the procedure described above and multiply them by the state or main unitary operator. The result is a random instance from a distribution that has a reduced bias. We emphasize here that it is not necessary for our purposes to use distributions that are uniform in the Haar measure. We only need to avoid special cases, e.g.~separable quantum states. Our approach, which is based on using random number generators almost everywhere, is essentially guaranteed to avoid such special cases.

\subsection{Quantum circuit structure}

Once we have generated a random instance of the target, i.e.~either a target state or a target unitary operator, we try to determine the minimum number of CNOT gates needed to implement the desired task. Considering a multi-qubit system (with the number of qubits $n>2$) and $N$ CNOT gates, there are multiple different possible configurations for the CNOT gates. For example, the quantum circuit below shows one example of a gate sequence containing three CZ gates applied to a four-qubit system.

\centerline{
\Qcircuit @C=1em @R=.7em {
& \gate{R} & \ctrl{2} & \gate{R} & \qw & \qw & \qw & \qw & \qw \\
& \gate{R} & \qw & \qw & \qw & \qw & \ctrl{1} & \gate{R} & \qw \\
& \gate{R} & \ctrl{0} & \gate{R} & \ctrl{1} & \gate{R} & \ctrl{0} & \gate{R} & \qw \\
& \gate{R} & \qw & \qw & \ctrl{0} & \gate{R} & \qw & \qw & \qw
}
}

\

\noindent A few points should be noted about the quantum circuit above. Firstly, the symbol $R$ denotes a single-qubit unitary operator but not a specific one. In other words, the $R$ operations in the quantum circuit above can all be different from each other. Secondly, we have not assigned a control and target qubit for each CZ gate. The reason is that the CZ gate is symmetric with respect to the two qubits. The CZ gate can be turned into a CNOT gate or vice versa by applying single-qubit gates before and after the two-qubit gate. Furthermore, the roles of the control and target qubits in a CNOT gate can be reversed by changing the single-qubit rotations applied before and after the CNOT gate, making it irrelevant which qubit is the control qubit and which qubit is the target qubit for purposes of finding the shortest gate sequence. Thirdly, apart from the single-qubit rotations applied to all the qubits in the first step of the quantum circuit, we apply single-qubit rotations only after each CZ gate. The reason is that a sequence of single-qubit rotations applied in succession to the same qubit can be combined into a single rotation. This property helps reduce the computation cost of our numerical calculations, in addition to being crucial for establishing the lower bound on the number of CNOT gates, as explained above.

The number of possible configurations for a two-qubit CZ gate in an $n$-qubit system is $n\times (n-1)/2$, which means that the total number of configurations in an $N$-gate sequence is $(n\times (n-1)/2)^N$. For a given target task, we considered all the possible gate configurations and used optimal-control-theory methods to find the single-qubit rotation parameters that give the maximum achievable fidelity for that gate configuration. The maximum achievable fidelity for a given $N$ is then the maximum fidelity obtained among all the different calculations. The shortest gate sequence needed for a perfect implementation of the task is the minimum value of $N$ that gives a fidelity value $F=1$, up to the unavoidable numerical errors.

We pause for a moment to comment on the reason for analyzing all possible gate configurations. Numerical optimization algorithms search the space of possible solutions to find the one solution that maximizes some objective, which in this case is the fidelity of the obtained state or unitary operator relative to the target state or operator. When the search space is continuous, one can use gradient-based search algorithms, which start from some initial guess for the solution and gradually move in the search space in the direction that maximizes the increase in fidelity. After a large number of iterations, and barring complications in the topography of the fidelity as a function of  the solution parameters, the algorithm converges to the optimal solution. This approach can be used for the single-qubit rotations in our problem, since these rotations can be expressed in terms of rotation angles, which form a continuous space. Finding the optimal CNOT gate configuration is trickier. The different configurations do not form a continuous space. As such, we cannot use standard gradient-based methods in the search for the optimal gate configuration. Furthermore, a simple change in one CNOT gate in the quantum circuit can have a large effect on the fidelity. We therefore use the brute-force approach in which we try every single one of the possible configurations and, by varying the parameters of the single-qubit rotations, determine the maximum achievable fidelity for every possible gate configuration. We thus acquire a list of CNOT gate configurations and their maximum achievable fidelities. We then select the gate configuration that corresponds to the highest fidelity.

\subsection{Numerical optimization algorithm}

For the optimization of the single-qubit rotations, we use a modified version of the gradient ascent pulse engineering (GRAPE) algorithm \cite{Khaneja}. In the standard GRAPE algorithm, the problem is formulated as a control problem where some parameters in a time-dependent Hamiltonian are varied in time to effect the desired operation. It is then assumed (typically as an approximation) that the system is controlled by piecewise constant pulses. In other words, the total pulse time is divided into $N$ time steps during which the Hamiltonian remains constant. The unitary evolution operator $U(T)$ of the dynamics can therefore be expressed as
\begin{equation}
U(T) = U_N U_{N-1} \cdots U_2 U_1,
\end{equation}
where $U_j$ is a unitary operator that describes the evolution in the $j$th time step:
\begin{equation}
U_j = \exp \left\{ -i \Delta t \left( \hat{H}_0 + \sum_{k=1}^{m} u_k(j) \hat{H}_k \right) \right\}.
\end{equation}
Here $\Delta t$ is the duration of the time step, $H_0$ is a time-independent term in the Hamiltonian, $m$ is the number of control parameters, $u_k(j)$ is the value of the $k$th control parameter in the $j$th time step, and $H_k$ is the $k$th control Hamiltonian. The algorithm proceeds by evaluating the derivative of the fidelity $F$ with respect to variations in all the control parameters $u_k(j)$, identifying the direction of the gradient and making a small move in the direction of the gradient, i.e.~adding a small correction to each $u_k(j)$ that is proportional to the derivative $\partial F / \partial u_k(j)$. For a sufficiently small step size, the update will increase the fidelity. After a large number of iterations, the fidelity is expected to approach its maximum achievable value for the situation under consideration. Importantly, when the time steps $\Delta t$ are small, the derivative $\partial F / \partial u_k(j)$ can be approximated by simple first-order expressions: for state preparation, the fidelity is defined as
\begin{equation}
F = {\rm Tr} \left\{ \rho_F U(T) \rho_0 U^{\dagger}(T) \right\},
\end{equation}
and its derivative is given by
\begin{equation}
\frac{\partial F}{\partial u_k(j)} = -i \Delta t \Big\langle \lambda_j \Big| \left[ \hat{H}_k, \rho_j \right] \Big\rangle,
\end{equation}
where $\rho_j$ is the density matrix propagated forward from the initial density matrix $\rho_0$, and $\lambda_j$ is the density matrix propagated backward from the target density matrix $\rho_F$. For unitary-operator synthesis, the fidelity is defined as
\begin{equation}
F = \left| \frac{ {\rm Tr} \left\{ U_F^{\dagger} U(T) \right\} }{2^n} \right|^2,
\end{equation}
and its derivative is given by
\begin{equation}
\frac{\partial F}{\partial u_k(j)} = \frac{2 \Delta t}{4^n} {\rm Im} \left\{ \Big\langle P_j \Big| \hat{H}_k X_j \Big\rangle \Big\langle X_j \Big| P_j \Big\rangle \right\},
\end{equation}
where $X_j$ is the unitary evolution operator propagated forward from the identity matrix, and $P_j$ is the unitary evolution operator propagated backward from the target unitary matrix $U_F$. The above expressions for $\partial F / \partial u_k(j)$ are derived in Ref.~\cite{Khaneja}.

In the version of the algorithm adapted for this work, we seek to optimize single-qubit rotations rather than control fields in a Hamiltonian. Nevertheless, the necessary modifications turn out to be rather straightforward. For a quantum circuit with a given gate configuration, the unitary evolution operator is given by
\begin{equation}
U_{\rm Total} = R_N V_{N} R_{N-1} V_{N-1} \cdots V_{2} R_{1} V_{1} R_{0},
\end{equation}
where $R_j$ is the combination (i.e.~product) of single-qubit rotations applied at the $j$th layer in the quantum circuit, and $V_j$ is the $j$th CZ gate in the gate configuration under consideration. Each gate configuration is defined by the sequence $V_j$, which are therefore kept fixed in the GRAPE optimization, and only the single-qubit rotations in $R_j$ are optimized for a given configuration. We update the single-qubit rotations by making a small detour: we first imagine that we can apply an additional single-qubit rotation after each single-qubit rotation in the quantum circuit. The reason for introducing separate single-qubit rotations, even though each one of the additional rotations follows a single-qubit rotation on the same qubit, is that the GRAPE algorithm uses a first-order approximation that is valid only when the unitary operators being updated are all close to the identity matrix. Meanwhile, the single-qubit rotations that are being updated and optimized in our calculations can be very far from the identity matrix. Introducing the small update rotations allows us to apply the GRAPE algorithm to these small rotations and treat them as variables whose parameters should be chosen to maximize the fidelity improvement. A small single-qubit rotation on qubit $k$ can be expressed as
\begin{equation}
R(\delta x_k, \delta y_k, \delta z_k) = \exp \left\{-i \left( \delta x_k \hat{\sigma}_x^{(k)} + \delta y_k \hat{\sigma}_y^{(k)} + \delta z_k \hat{\sigma}_z^{(k)} \right) \right\},
\end{equation}
where $\hat{\sigma}_x^{(k)}$, $\hat{\sigma}_y^{(k)}$ and $\hat{\sigma}_z^{(k)}$ are the standard Pauli matrices for qubit $k$. The three parameters $\delta x_k$, $\delta y_k$ and $\delta z_k$ can then be treated as the $u_k(j)$ parameters in the GRAPE algorithm, with corresponding control Hamiltonians (i.e.~$\hat{H}_k$) $\hat{\sigma}_x^{(k)}$, $\hat{\sigma}_y^{(k)}$ and $\hat{\sigma}_z^{(k)}$. Once the derivatives and gradient of $F$ are evaluated, each small rotation $R(\delta x_k, \delta y_k, \delta z_k)$ is determined and multiplied by the single-qubit rotation preceding it. As a result, the single-qubit rotations in the gate sequence are updated and the fidelity is slightly increased. The process is repeated a large number of iterations to obtain the maximum achievable fidelity. It is worth mentioning here that our approach can be thought of as taking each step in the quantum circuit and treating it as a time step in the GRAPE algorithm, although there is no time variable in the gate decomposition problem (apart from the time ordering of the single- and two-qubit gates in the quantum circuit). In other words, our calculations give the optimal single-qubit rotations but not the time needed to implement these rotations in a certain physical system.

As explained above, we systematically go over all possible gate configurations for a given number of CNOT gates. For each configuration, we start by randomly generating single qubit rotations. Then we use the GRAPE algorithm to update these rotations. As a general rule, we run $10^3$ optimization iterations for each situation under consideration, and we add more iterations if we determine that it is needed to reduce numerical errors or if there are discernible fluctuations in the results indicating slow convergence.

\section{Numerical results}
\label{Sec:Results}

We now present the results of our numerical calculations. We start with the results for arbitrary targets. Then we present results for synthesizing Toffoli gates.

\subsection{State preparation}

\begin{figure}[h]
\includegraphics[width=8.0cm]{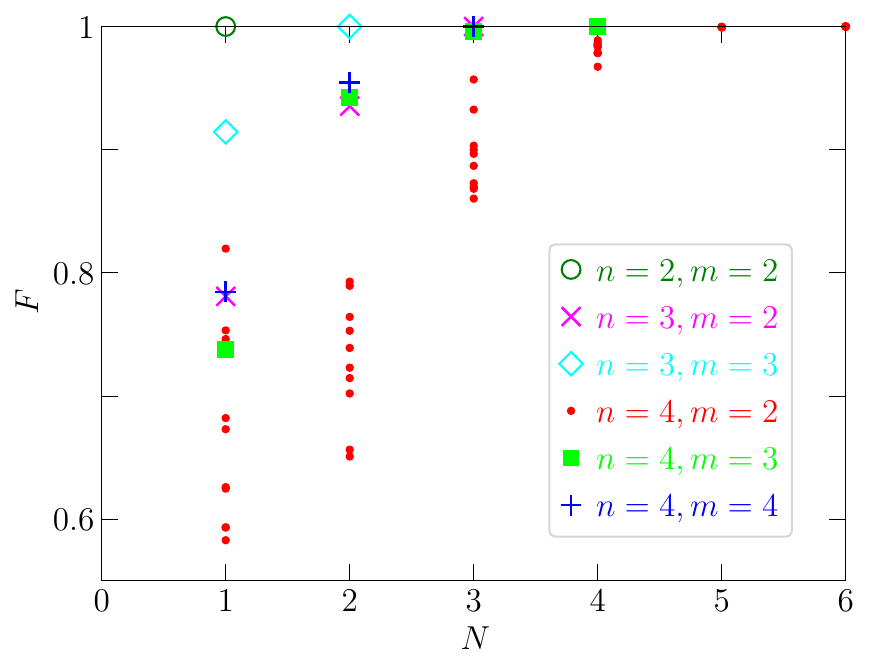}
\caption{The maximum achievable fidelity $F$ for $n$-qubit state preparation when using a quantum circuit that contains $N$ entangling gates. The green open circle corresponds to preparing a two-qubit state (i.e.~$n=2$) using two-qubit CZ gates (i.e.~$m=2$). The magenta $\times$ symbols and cyan diamonds are for the case $n=3$ and correspond, respectively, to using two- and three-qubit CZ gates. The red dots, green squares and blue + symbols are for the case $n=4$ and correspond, respectively, to using two-, three- and four-qubit CZ gates. We used ten different randomly generated target states for each setting. To show the statistical spread in the results, the red dots show the results for all ten instances of the target state for $n=4$ and $m=2$. For all other data points, we took the lowest value of maximum fidelity among the ten random instances. For a two-qubit system, a single CZ gate is sufficient for perfect state preparation. For all other data sets, the last shown (i.e.~largest-$N$) data point in each set has an infidelity $1-F \sim 10^{-7}$ or less after $10^3$ optimization iterations and shrinks to below $10^{-12}$ if we continue the optimization to $10^6$ iterations, which implies that the small numbers that we obtain for $1-F$ are numerical errors, and the actual achievable fidelity in each one of these cases is $F=1$.}
\label{Fig:FidelityStatePreparation}
\end{figure}

In all our calculations for state preparation, the initial state is the product state in which each qubit is in the state $\ket{0}$. For each system size and elementary gate set, we plot the fidelity $F$ as a function of the number of entangling gates $N$. The fidelity increases as a function of $N$ until it reaches $F=1$ for a certain value of $N$, which we can identify as the minimum number of entangling gates needed for perfect state preparation.

For a two-qubit system, the calculation is simple. There is only one possible configuration for the CZ gates. Our numerical calculations reproduce the well-known result that a single CZ (or CNOT) gate is sufficient to prepare an arbitrary target state \cite{ShendeStatePrep}. This result is represented by the green open circle in Fig.~\ref{Fig:FidelityStatePreparation}. For a three-qubit system and two-qubit CNOT gates, each CNOT gate has three possible configurations corresponding to the three possible pairings of the three qubits. The total number of possible configurations is therefore $3^N$. The fidelity $F$ reaches 1 at $N=3$, which means that with three CNOT gates we can prepare an arbitrary target state. Specifically, at $N=3$ we obtain $F=1$ up to numerical errors (on the order of $10^{-12}$). This result is in agreement with the protocol in \cite{Znidaric} and is higher than the lower bound ($N=2$) based on the parameter-counting calculation in Sec.~\ref{Sec:Background}. If for the entangling gates we use three-qubit CZ gates (cyan diamonds in Fig.~\ref{Fig:FidelityStatePreparation}), two entangling gates are needed for perfect state preparation. If we use the generalized parameter-counting formulae given in Sec.~\ref{Sec:Background}, we find that the lower bound for three-qubit state preparation using three-qubit CZ gates is $N=2$. Hence our numerical results in this case agree with the lower bound.

Next we consider the case of preparing a four-qubit state using two-qubit CNOT gates, in addition to single-qubit rotations. The red dots in Fig.~\ref{Fig:FidelityStatePreparation} show the results for ten different target states for each quantum circuit size. As could be expected, there are large variations in the fidelity for short quantum circuits, as the randomly generated target states could be close or far from states that can be reached with a small number of CNOT gates. The lowest dot can be considered a good conservative estimate for the fidelity for the hardest-to-reach target states. As the quantum circuit size increases, the fidelity increases and instance-to-instance variations decrease. For a quantum circuit with six CNOT gates (i.e.~$N=6$), any target state can be prepared perfectly, i.e.~with fidelity $F=1$. Specifically, the numerical results give $F$ values that keep approaching 1 until the minimum infidelity $1-F \lesssim 10^{-12}$. This result suggests that the asymptotic value of $1-F$ is exactly zero, up to numerical rounding errors. In contrast, for $N=5$ the minimum infidelity converges to $1-F\sim 10^{-4}$ even if we require that $F$ converges at the level of $10^{-12}$. We therefore conclude that the fidelity values obtained in this case are accurate values for the fidelity that cannot be exceeded with $N=5$. In other words, perfect state preparation is not possible with $N=5$. The number $N=6$ coincides with the theoretical lower bound and is shorter than the size ($N=9$) of the gate sequence proposed in Ref.~\cite{Plesch}. For the case of preparing a four-qubit state using multi-qubit CZ gates, the lower bounds are $N=4$ for three-qubit CZ gates and $N=3$ for four-qubit CZ gates. Our numerical results show that the minimum circuit sizes in these cases are $N=4$ and $N=3$, in agreement with the respective lower bounds.

It is interesting that even below the lower bound for perfect state preparation, the fidelity can be remarkably high. For example, all ten red dots at $N=5$ in Fig.~\ref{Fig:FidelityStatePreparation} lie in the range 0.9993-0.9997. In a realistic setup, especially in the near future, the increase of $\sim 10^{-4}$ in fidelity that we gain in going from $N=5$ to $N=6$ could be offset by errors that are introduced by the extra CNOT gate in the quantum circuit. In such a situation, it can be optimal to use a quantum circuit with five CNOT gates, rather than the six-CNOT-gate circuit based on the lower bound for perfect state preparation.

Because we have fidelity values for all the different gate configurations, we can go beyond identifying a single optimal control sequence and also analyze the statistics of how well different gate configurations perform. Before doing that, however, we consider a geometric analogy between state preparation and a point moving in a multidimensional space. If we want to reach an arbitrary point in three dimensional Euclidean space, we can first move in the x direction, then move in the y direction and finally move in the z direction. Any permutation of these three steps allows access to any point in the whole space. We can similarly expect that if one gate configuration allows a perfect preparation of a random target state any alternative configuration that is obtained by a permutation of the qubit labels will also allow a perfect preparation of the state. We examine this point by focusing on one of our ten target states first. To make sure that we do not have an accidentally easy random instance, we took the target state that corresponds to the lowest red dot at $N=4$ in Fig.~\ref{Fig:FidelityStatePreparation} and used it as the target state. We then performed the optimization algorithm with $N=6$. The maximum fidelity that we obtained in this case reached $1-F\sim 10^{-12}$. We took the corresponding quantum circuit and inspected the fidelity data for the 24 permutations obtained by qubit relabeling. All permutations gave $F=1$ (up to numerical errors) after a sufficient number of optimization iterations \cite{ConvergenceFootnote}. Similarly, with a few random checks, we verified and confirmed that the configurations that give $F=1$ for any one of our ten randomly generated target states give $F=1$ for all target states. This result also supports the idea that there is a geometric reason that makes a certain gate configuration able to reach any target state in the $n$-qubit Hilbert space.

\begin{figure}[h]
\includegraphics[width=8.0cm]{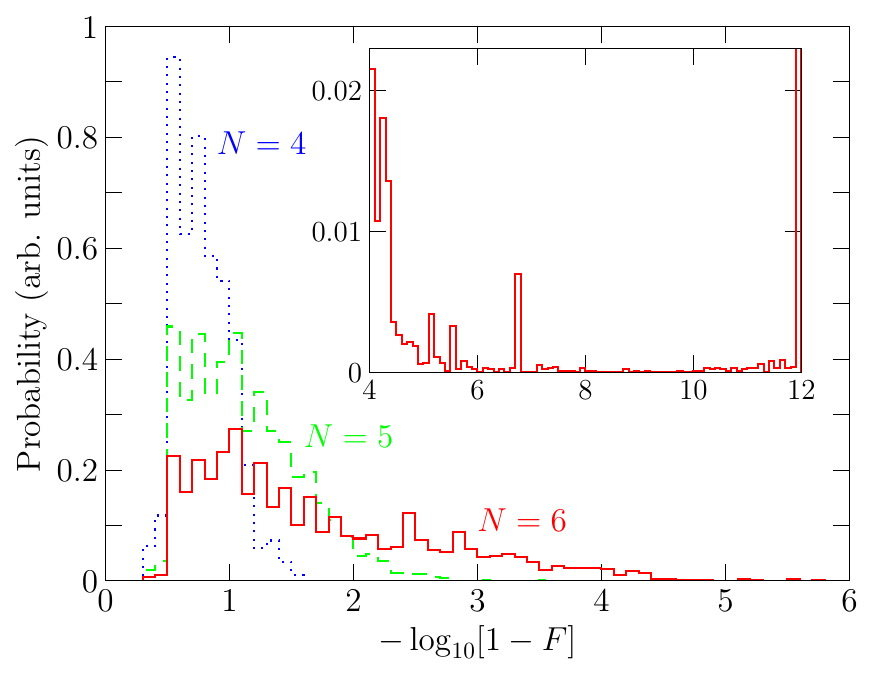}
\caption{Histograms of fidelity $F$ values for all the possible CNOT gate configurations for one instance of the state-preparation problem with a four-qubit target state and using two-qubit CNOT gates as the entangling gates. A logarithm function is used for the x axis to magnify the region close to $F=1$ and make the features of the histogram in this region easier to discern. The dotted (blue), dashed (green) and solid (red) lines correspond, respectively, to $N=4$, 5 and 6. The inset shows the high-fidelity tail of the $N=6$ histogram. The peak at $1-F=10^{-12}$ includes all the configurations that gave $1-F<10^{-12}$, which turn out to be 20\% of all possible configurations.}
\label{Fig:HistogramStatePrep}
\end{figure}

We now look at the statistics of gate fidelity data for all different gate configurations for the cases $N=4$, 5, and 6. It is worth noting here that the number of configurations for $N=6$ is $6^6=46,656$. The histograms in Fig.~\ref{Fig:HistogramStatePrep} show the number of configurations that have any given value of $F$, all for the same target state. As above, we use the target state that gave the lowest value of $F$ at $N=4$ in Fig.~\ref{Fig:FidelityStatePreparation}; we note, however, that all the instances whose histograms we inspected gave qualitatively similar results. Instead of using the fidelity $F$ as the x axis, we use the function $-\log_{10}[1-F]$, which magnifies the region just below $F=1$. Up to $N=5$ the fidelity data form relatively simple distributions that end abruptly at some finite values of $F$ that depend on $N$ and are all smaller than 1. For $N=6$, the histogram has a long high-fidelity tail with a high peak at $1-F\sim 10^{-12}$. In theory, the configurations that allow perfect state preparation should give $F$ values that keep moving closer to $F=1$ if we continue the optimization indefinitely. In practice, however, numerical rounding errors stop this trend and make the optimization procedure unreliable when $1-F\sim 10^{-12}$. We therefore terminate the optimization if we obtain $1-F<10^{-12}$.

One of our main goals in plotting these histograms is to identify the number of gate configurations that allow a perfect state preparation. For this purpose, it would be ideal if for $N=6$ the tail ended at some value of $F$ (e.g.~a value with $1-F\sim 10^{-5}$) and were then followed by a well-separated peak at $1-F\lesssim 10^{-12}$ that includes all the $F=1$ configurations. This shape of histogram could be expected based on the intuitive picture of the gate configurations corresponding to a discrete variable: unlike continuous variables, unless a certain gate configuration allows perfect state preparation (i.e.~$F=1$), it should not allow us to approach the target state by a distance that corresponds to, say, $1-F\sim 10^{-10}$, which seems too small for a randomly generated state with the small numbers $n=4$ and $N=6$. While the histogram exhibits a relatively sharp drop at $-\log_{10}[1-F]\approx 4.5$, a small peak appears around $-\log_{10}[1-F]\approx 7$ and a small tail persists up to the peak at $-\log_{10}[1-F]=12$. We suspect that all of the data points with $-\log_{10}[1-F]>7$ correspond to $F=1$ but the optimization algorithm incorrectly identified them as having converged at lower values. The peak at $1-F<10^{-12}$, which we can confidently identify as corresponding to $F=1$, contains 8611 data points. Hence the number of different gate configurations in the $F=1$ peak is at least 8611. If we also include gate configurations that are obtained by qubit permutations of $F=1$ configurations, the total number of $F=1$ configurations rises slightly and becomes 9264, which is about 20\% of all possible gate configurations. In other words, there are a remarkably large number of gate configurations that allow a perfect preparation of an arbitrary target state. Although some multiplicity is to be expected based on qubit relabeling and commuting gate configurations, there are still a large number of qualitatively dissimilar gate configurations. For example, the following three quantum circuits gave the highest values of $F$ after $10^4$ iterations in the first run of the algorithm (all of which gave $1-F<10^{-11}$ and can hence be confidently identified as having $F=1$):

\centerline{
\Qcircuit @C=1em @R=.7em {
& \gate{R} & \ctrl{1} & \gate{R} & \ctrl{3} & \gate{R} & \ctrl{2} & \gate{R} & \ctrl{3} & \gate{R} & \qw & \qw & \qw & \qw & \qw \\
& \gate{R} & \ctrl{0} & \gate{R} & \qw & \qw & \qw & \qw & \qw & \qw & \ctrl{1} & \gate{R} & \qw & \qw & \qw \\
& \gate{R} & \qw & \qw & \qw & \qw & \ctrl{0} & \gate{R} & \qw & \qw & \ctrl{0} & \gate{R} & \ctrl{1} & \gate{R} & \qw \\
& \gate{R} & \qw & \qw & \ctrl{0} & \gate{R} & \qw & \qw & \ctrl{0} & \gate{R} & \qw & \qw & \ctrl{0} & \gate{R} & \qw
\gategroup{1}{9}{4}{10}{0.7em}{--}
\gategroup{2}{11}{3}{12}{0.7em}{--}
}
}

\

\

\centerline{
\Qcircuit @C=1em @R=.7em {
& \gate{R} & \qw & \qw & \qw & \qw & \ctrl{1} & \gate{R} & \ctrl{2} & \gate{R} & \qw & \qw & \ctrl{3} & \gate{R} & \qw \\
& \gate{R} & \qw & \qw & \ctrl{1} & \gate{R} & \ctrl{0} & \gate{R} & \qw & \qw & \qw & \qw & \qw & \qw & \qw \\
& \gate{R} & \ctrl{1} & \gate{R} & \ctrl{0} & \qw & \qw & \qw & \ctrl{0} & \gate{R} & \ctrl{1} & \gate{R} & \qw & \qw & \qw \\
& \gate{R} & \ctrl{0} & \gate{R} & \qw & \qw & \qw & \qw & \qw & \qw & \ctrl{0} & \gate{R} & \ctrl{0} & \gate{R} & \qw
}
}

\

\

\centerline{
\Qcircuit @C=1em @R=.7em {
& \gate{R} & \qw & \qw & \qw & \qw & \ctrl{2} & \gate{R} & \ctrl{3} & \gate{R} & \ctrl{1} & \gate{R} & \ctrl{2} & \gate{R} & \qw \\
& \gate{R} & \qw & \qw & \ctrl{1} & \gate{R} & \qw & \qw & \qw & \qw & \ctrl{0} & \gate{R} & \qw & \qw & \qw \\
& \gate{R} & \ctrl{1} & \gate{R} & \ctrl{0} & \gate{R} & \ctrl{0} & \gate{R} & \qw & \qw & \qw & \qw & \ctrl{0} & \gate{R} & \qw \\
& \gate{R} & \ctrl{0} & \gate{R} & \qw & \qw & \qw & \qw & \ctrl{0} & \gate{R} & \qw & \qw & \qw & \qw & \qw
}
}

\

\noindent Each one of these quantum circuits can be used to generate 23 other equivalent ones by qubit permutations. Furthermore, the fourth and fifth CZ gates in the first quantum circuit operate on different qubits and therefore commute with each other, which generates another 24 quantum circuits that must also give $F=1$. However, the three quantum circuits shown above do not seem to be easily convertible into each other via qubit permutation or CZ gate commutation. As an indication that the third quantum circuit cannot be transformed into either of the other two by simple permutations and commutation, we note that the first and third qubits are involved in four CZ gates while the second and fourth quits are involved in two of CZ gates. In the other two quantum circuits, only one of the qubits is involved in four CNOT gates.

\begin{figure}[h]
\includegraphics[width=8.0cm]{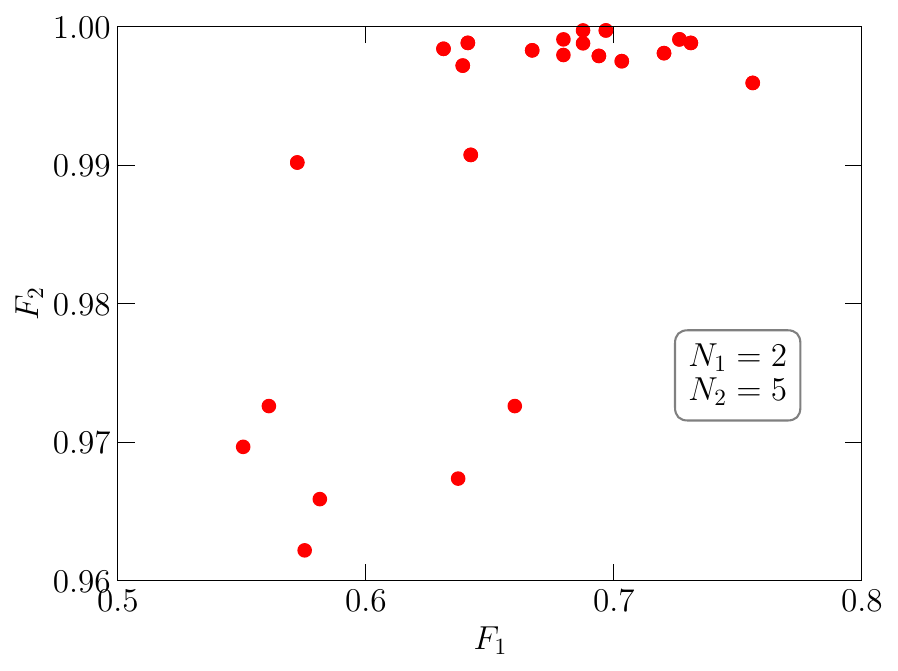}
\caption{Correlations between fidelity values for quantum circuits of different sizes, specifically quantum circuits of size $N_1=2$ and those of size $N_2=5$. Each gate configuration (to which we refer as $Q_i$) in the set of all configurations with size $N_1$ produces one point in the plot. The x axis shows the fidelity $F_1$ for the quantum circuit $Q_i$. The y axis shows the maximum fidelity $F_2$ that can be achieved with a quantum circuit of size $N_2$ and containing $Q_i$ as its initial part. While there is some correlation between $F_1$ and $F_2$, the highest value of $F_2$ does not correspond to the highest value of $F_1$, indicating that optimizing the quantum circuit one layer at a time does not produce the global optimum among large quantum circuits.}
\label{Fig:Correlations}
\end{figure}

It is also interesting to look at correlations between fidelity values for quantum circuits of different sizes. These correlations address the question of whether one could start by optimizing a quantum circuit with one CNOT gate, then add one layer (i.e.~another CNOT gate and the corresponding single-qubit rotations) and optimize the parameters of the newly added layer while keeping the already-optimized parameters from the first layer fixed, and so on. If this approach produces the maximum value of $F$ for large quantum circuits, it would greatly speed up the optimization of large quantum circuits. In Fig.~\ref{Fig:Correlations} we plot the maximum achievable fidelity $F_2$ of a quantum circuit of size $N_2=5$ as a function of the fidelity $F_1$ that is achievable with the initial part of the quantum circuit, specifically the first two layers of the circuit. Not surprisingly, there is some correlation between $F_1$ and $F_2$, especially for small values of fidelity. These correlations can be understood based on the intuitive idea that bad small quantum circuits (e.g.~those with the same CNOT gate repeated multiple times in a row) tend to lead to bad larger circuits. A more important result in Fig.~\ref{Fig:Correlations} is the fact that the highest values of $F_2$ do not correspond to the highest values of $F_1$. This result means that, for example, the first two layers in the optimal five-layer quantum circuit are not the optimal two-layer circuit. In other words, if we construct the quantum circuit by gradually increasing the circuit size and optimizing each layer when it is added to the circuit, we will in general not obtain the optimal quantum circuit. The optimization must be performed globally to ensure finding the optimal circuit of a given size.

\subsection{Quantum circuit depth}

An important question when designing quantum circuits is the depth of the circuit. For example, the first of the three quantum circuits shown in the previous subsection has a smaller depth and can be implemented faster than the other two, because the two steps enclosed by the dashed lines can be implemented simultaneously. We took all 9264 gate configurations that gave $F=1$ and calculated the depth for each one of them. The depths of the quantum circuits had the distribution \{0, 0, 0, 1008, 3984, 4272\}. In other words, the minimum depth is 4, with 1008 different quantum circuits having that minimum depth. One example of a minimum-depth quantum circuit is:

\centerline{
\Qcircuit @C=1em @R=.7em {
& \gate{R} & \ctrl{1} & \gate{R} & \ctrl{2} & \gate{R} & \qw & \push{\rule{0em}{1.2em}} \qw & \ctrl{1} & \gate{R} & \ctrl{1} & \gate{R} & \qw & \push{\rule{0em}{1.2em}} \qw & \qw \\
& \gate{R} & \ctrl{0} & \gate{R} & \qw & \qw & \ctrl{2} & \gate{R} & \ctrl{0} & \gate{R} & \ctrl{0} & \gate{R} & \qw & \qw & \qw \\
& \gate{R} & \qw & \qw & \ctrl{0} & \gate{R} & \qw & \qw & \qw & \qw & \qw & \qw & \ctrl{1} & \gate{R} & \qw \\
& \gate{R} & \qw & \qw & \qw & \qw & \ctrl{0} & \gate{R} & \qw & \qw & \qw & \qw & \ctrl{0} & \gate{R} & \qw
\gategroup{1}{5}{4}{8}{0.7em}{--}
\gategroup{1}{11}{4}{14}{0.7em}{--}
%
%
%
%
}
}

\

\noindent The two pairs of steps that can each be parallellized are shown by the dashed boxes.

The quantum circuit shown above reveals another interesting feature: not all qubit pairs appear in the CNOT gate sequence. In particular, there are no CNOT gates on the qubit pair 1-4 or the pair 2-3. This property could be helpful in the design or utilization of real devices. For example, if a certain multi-qubit device realized in experiment has one defective coupling between a pair of qubits, we can look for quantum circuits that do not utilize that particular qubit pair. The fact that there are a large number of alternative gate sequences allows us to look for the one that is optimal for implementation on the experimental device under consideration. Although our computational limitations allow us to establish this result only for the case of four-qubit quantum state preparation, it seems likely that a similar situation will arise for larger systems and/or for unitary operator synthesis.

\subsection{Unitary operator synthesis - arbitrary target}

\begin{figure}[h]
\includegraphics[width=8.0cm]{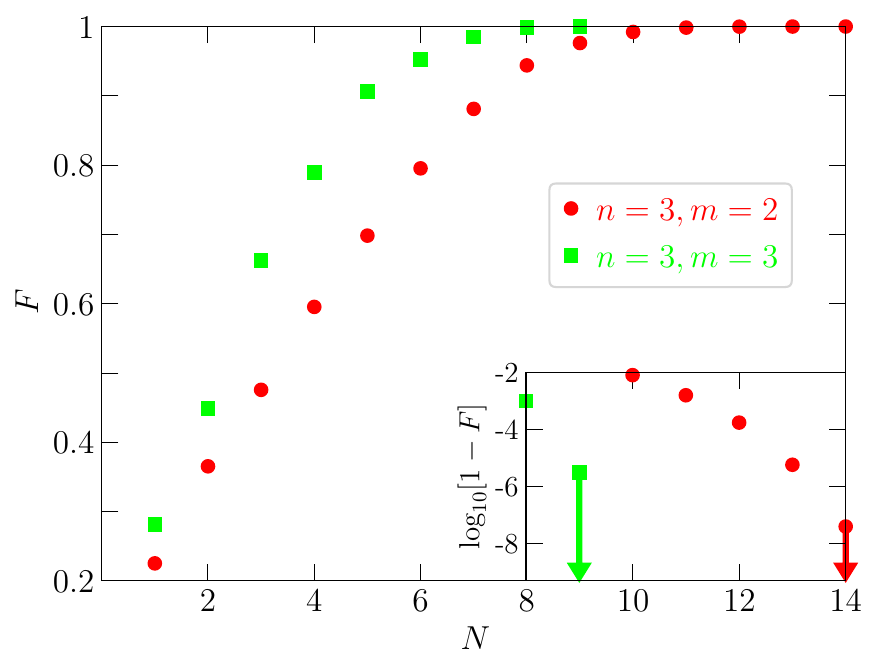}
\caption{The maximum achievable fidelity $F$ for three-qubit unitary operator synthesis as a function of quantum circuit size $N$. The red circles and green squares correspond, respectively, to using two- and three-qubit CZ gates. Randomly generated unitary operators were used as target operators. The inset shows the logarithm of the infidelity $\log_{10}[1-F]$ for the high-fidelity points. The arrows indicate that $F$ can be made arbitrarily close to $F=1$ for the last point in each data set by increasing the number of optimization iterations, while other data points do not experience significant changes with an increased number of iterations.}
\label{Fig:FidelityUnitary}
\end{figure}

Next we turn to the case of decomposing, or synthesizing, $n$-qubit unitary operators. Figure \ref{Fig:FidelityUnitary} shows the results for arbitrary three-qubit unitary operators decomposed into elementary gates. The fidelity reaches $F=1$ at $N=14$, in agreement with the lower bound and slightly shorter than the size ($N=15$) of the decomposition proposed in Ref.~\cite{Rakyta}. It is interesting to note that the fidelity goes above 0.99 already for $N=10$. As a result, extremely high fidelities can be obtained even with circuit sizes that are smaller than the perfect-decomposition lower bound. This result provides concrete quantitative benchmarks for the approximate quantum circuit synthesis of three-qubit unitary operators \cite{Camps}. We note that in the unitary operator synthesis calculations, we used ten different instances up to $N=10$ and used fewer instances for larger values of $N$, because the computation time became significant. For the case $N=14$, the calculation took the equivalent of a few months on a single core of a present-day computer. For all values of $N$, the instance-to-instance fluctuations were small and suggested that even with small numbers of instances, we can expect the numerical results to accurately represent the statistical average. We also performed calculations where we used the three-qubit CZ gate (which is equivalent to the three-qubit Toffoli gate) instead of two-qubit CNOT gates in the elementary gate set. The minimum circuit size needed to perfectly reproduce an arbitrary three-qubit unitary operator was $N=9$ in this case. This result agrees with the respective lower bound.

\begin{figure}[h]
\includegraphics[width=8.0cm]{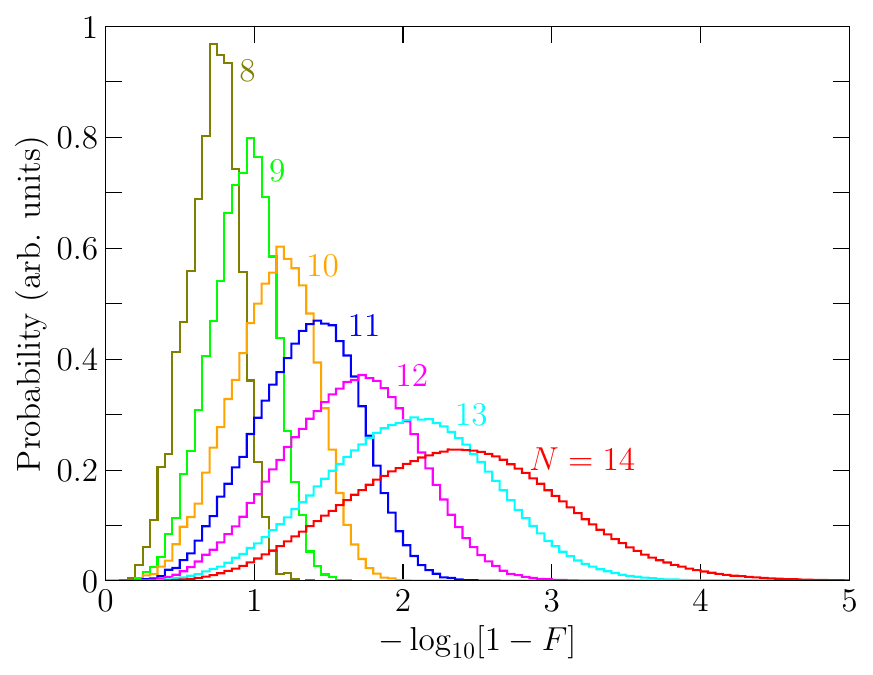}
\caption{Histograms of fidelity $F$ values for all the possible CNOT gate configurations for the problem of synthesizing a three-qubit unitary operator using two-qubit CNOT gates and single-qubit gates. The olive, green, orange, blue, magenta, cyan and red lines correspond, respectively, to $N=8$, 9, 10, 11, 12, 13 and 14. All the data in this figure were obtained using $10^3$ optimization iterations.}
\label{Fig:HistogramUnitary}
\end{figure}

Figure \ref{Fig:HistogramUnitary} shows histograms of the fidelity data for different quantum circuit sizes. It is worth noting that these histograms do not show any qualitative change between $N=14$, where perfect synthesis becomes possible, and smaller values of $N$. However, although it is not visible at the y-axis scale used in Fig.~\ref{Fig:HistogramUnitary}, the $N=14$ data has a long tail extending to higher fidelities. After $10^3$ optimization iterations, the highest fidelities that we obtained in that case had $1-F\sim 10^{-8}$. Out of the $3^{14}=4,782,969$ different configurations, 5 configurations had $1-F<10^{-7}$, and 3553 configurations had $1-F<10^{-5}$. We took the top five configurations and inspected the fidelity values for their permutations. Most permutations gave $10^{-4}<1-F<10^{-3}$. We reran the algorithm on all of them and confirmed that with a few randomly chosen initial guess choices and $10^4$ iterations all permutations give $1-F<10^{-8}$. We then reran the algorithm with $10^4$ iterations on all gate configurations that gave $1-F<10^{-3}$. If we identify any gate sequence with a numerical fidelity value of $1-F<10^{-8}$ and all its permutations as having $F=1$, we find that about $9.1 \times 10^{5}$ gate sequences, i.e.~about 20\% of all possible sequences, meet these criteria. As in the case of state preparation, a remarkably large number of distinct quantum circuits allow a perfect synthesis of arbitrary unitary operators, even at the lower bound for the quantum circuit size. The minimum depth of the quantum circuit needed for perfect unitary operator synthesis must be the same as the number of CNOT gates, i.e.~$N=14$, because in a three-qubit system it is not possible to have two consecutive two-qubit gates with no overlap in the qubit pairs involved in the two gates.

\begin{figure}[h]
\includegraphics[width=8.0cm]{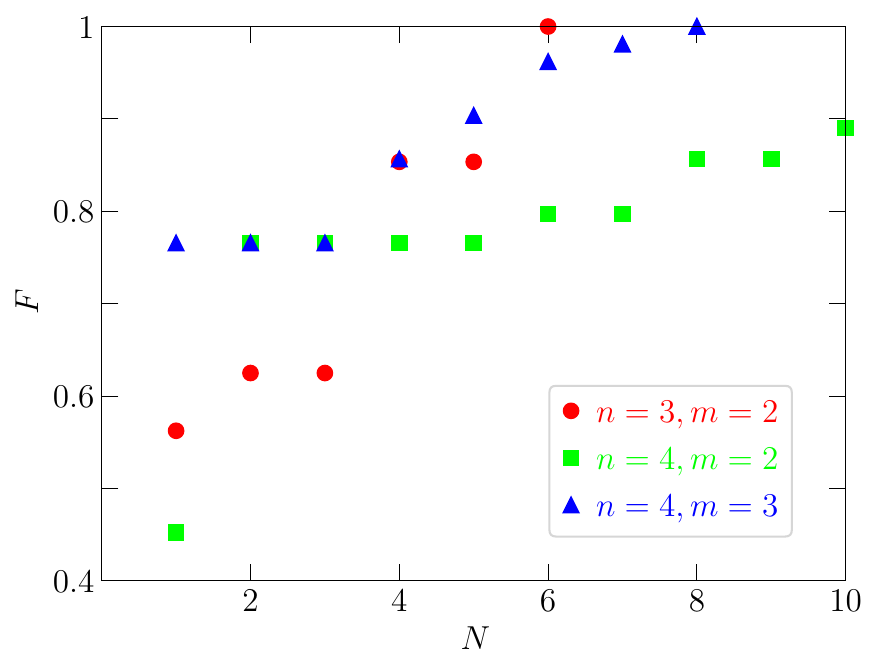}
\caption{The maximum achievable fidelity $F$ as a function of quantum circuit size $N$ for synthesizing a multi-qubit Toffoli gate. The red circles are for the well-known case of synthesizing a three-qubit Toffoli gate from two-qubit CNOT gates. The green squares are for the case of synthesizing a four-qubit Toffoli gate from two-qubit CNOT gates. The blue triangles are for the case of synthesizing a four-qubit Toffoli gate from three-qubit Toffoli gates.}
\label{Fig:FidelityToffoli}
\end{figure}

\subsection{Unitary operator synthesis - multi-qubit Toffoli gate}

In addition to using arbitrary unitary operators as the target gates, we also considered the special case of decomposing Toffoli gates into smaller elementary gates. It should be noted that such special cases can be decomposable into shorter circuits than those needed for the general case of an arbitrary (i.e.~worst-case) target. In fact, for the $n$-qubit Toffoli gate, it has been shown that the minimum number of CNOT gates needed for perfect synthesis grows at most quadratically with $n$ \cite{Barenco,Saeedi}. In other words, perfect synthesis of the $n$-qubit Toffoli gate is possible with $\sim n^2$ or fewer 2-qubit CNOT gates, in contrast to the exponential scaling for arbitrary unitary operators.

We start with the well-known case of decomposing the three-qubit Toffoli gate into two-qubit CNOT gates \cite{Nielsen}. The fidelity $F$ as a function of quantum circuit size $N$ is plotted in Fig.~\ref{Fig:FidelityToffoli}. The data shows that six CNOT gates are needed to perfectly synthesize the three-qubit Toffoli gate. Interestingly, in the plot of fidelity vs circuit size, two flat steps are encountered: $N=2$ and $N=3$ give the same value of $F$, and similarly $N=4$ and $N=5$ give the same value of $F$. We found that out of the $3^6=729$ possible gate configurations 54 different configurations give $F=1$. Since the three-qubit Toffoli gate is equivalent to the three-qubit CZ gate and the latter is symmetric with respect to permutations of the three qubits, a six-fold symmetry resulting from qubit permutations is expected. As a result, we find 54/6=9 dissimilar configurations. Furthermore, since the three-qubit Toffoli gate is its own inverse, any CNOT gate configuration that gives $F=1$ can be reversed in time to produce another configuration that also gives $F=1$. This consideration allows us to reduce the number of dissimilar configurations to 6. The CZ (or CNOT) gate configurations in these quantum circuits are:

\centerline{
\Qcircuit @C=1em @R=.7em {
& \ctrl{1} & \ctrl{1} & \ctrl{2} & \qw & \ctrl{2} & \qw & \qw \\
& \ctrl{0} & \ctrl{0} & \qw & \ctrl{1} & \qw & \ctrl{1} & \qw \\
& \qw & \qw & \ctrl{0} & \ctrl{0} & \ctrl{0} & \ctrl{0} & \qw
}
}

\

\centerline{
\Qcircuit @C=1em @R=.7em {
& \ctrl{1} & \ctrl{2} & \ctrl{1} & \qw & \ctrl{2} & \qw & \qw \\
& \ctrl{0} & \qw & \ctrl{0} & \ctrl{1} & \qw & \ctrl{1} & \qw \\
& \qw & \ctrl{0} & \qw & \ctrl{0} & \ctrl{0} & \ctrl{0} & \qw
}
}

\

\centerline{
\Qcircuit @C=1em @R=.7em {
& \ctrl{1} & \ctrl{2} & \ctrl{1} & \qw & \qw & \ctrl{2} & \qw \\
& \ctrl{0} & \qw & \ctrl{0} & \ctrl{1} & \ctrl{1} & \qw & \qw \\
& \qw & \ctrl{0} & \qw & \ctrl{0} & \ctrl{0} & \ctrl{0} & \qw
}
}

\

\centerline{
\Qcircuit @C=1em @R=.7em {
& \ctrl{1} & \ctrl{2} & \qw & \ctrl{1} & \qw & \ctrl{2} & \qw \\
& \ctrl{0} & \qw & \ctrl{1} & \ctrl{0} & \ctrl{1} & \qw & \qw \\
& \qw & \ctrl{0} & \ctrl{0} & \qw & \ctrl{0} & \ctrl{0} & \qw
}
}

\

\centerline{
\Qcircuit @C=1em @R=.7em {
& \ctrl{1} & \ctrl{2} & \qw & \ctrl{2} & \qw & \ctrl{1} & \qw \\
& \ctrl{0} & \qw & \ctrl{1} & \qw & \ctrl{1} & \ctrl{0} & \qw \\
& \qw & \ctrl{0} & \ctrl{0} & \ctrl{0} & \ctrl{0} & \qw & \qw
}
}

\

\centerline{
\Qcircuit @C=1em @R=.7em {
& \ctrl{1} & \ctrl{2} & \qw & \qw & \ctrl{1} & \ctrl{2} & \qw \\
& \ctrl{0} & \qw & \ctrl{1} & \ctrl{1} & \ctrl{0} & \qw & \qw \\
& \qw & \ctrl{0} & \ctrl{0} & \ctrl{0} & \qw & \ctrl{0} & \qw
}
}

\

\noindent In all the perfect decompositions, each one of the three qubits is involved in four CNOT gates, i.e.~each combination of qubit pairs appears twice in the CNOT gates of the quantum circuit. There were configurations in which the same CNOT gate was repeated twice in succession, and there were configurations in which all adjacent CNOT gates were different from each other, i.e.~involved different qubits. It is also worth noting here that there are a total of 90 different configurations of six-CNOT-gate sequences where each combination of qubit pairs appears twice in the sequence, and the majority of these (54/90=60\%) can be used to synthesize a perfect Toffoli gate.

For the case of decomposing the four-qubit Toffoli gate into two-qubit CNOT gates, we performed calculations for up to $N=10$. The case $N=10$ took the equivalent of a few months of single-core computation time, which means that it was too computationally costly to go beyond $N=10$. The fidelity reached only about 0.9 at $N=10$, which means that a few more CNOT gates are probably needed to obtain a perfect decomposition of the four-qubit Toffoli gate. We note that the general-case lower bound for an arbitrary $n=4$ unitary operator is $N=61$.

For the case of decomposing the four-qubit Toffoli gate into three-qubit Toffoli gates and single-qubit gates, we find that eight three-qubit Toffoli gates are needed for a perfect decomposition. In this calculation we used only three-qubit CZ gates and single-qubit rotations in the elementary gate set, i.e.~not including two-qubit CNOT gates. If we include both two- and three-qubit CZ gates in the elementary gate set, we do not obtain any increase in $F$ compared to the case where we use only three-qubit CZ gates. As a result, the minimum number of entangling gates is still $N=8$. It should be noted here that there is a well-known decomposition of the $n$-qubit Toffoli gate into two $(n-1)$-qubit Toffoli gates, two rotations controlled by a single qubit and one rotation controlled by $(n-2)$ qubits \cite{Barenco,Saeedi}. Each controlled rotation can be decomposed into two controlled NOT operations and single-qubit rotations. The gate count then becomes four $(n-1)$-qubit Toffoli gates and four CNOT gates, i.e.~a total of eight entangling gates. Our results show that, at least for the case $n=4$, there is no shorter quantum circuit that achieves the same goal of perfectly synthesizing the $n$-qubit Toffoli gate.

Finally, it is worth emphasizing that the minimum number of gates needed for a perfect decomposition can depend on the entangling gates in the elementary gate set. For example, if the entangling gate in the elementary gate set is the general controlled-U gate instead of just the CNOT gate, the three-qubit Toffoli gate can be decomposed into five two-qubit gates, in addition to single-qubit rotations \cite{Nielsen}. We performed numerical calculations of this case, treating the rotations U in the controlled-U gates as variables to be optimized, and we confirmed that the minimum number of gates when using controlled-U gates is five. This result shows that the five-gate decomposition in the literature is optimal in terms of quantum circuit size.

\section{Conclusion}
\label{Sec:Conclusion}

In conclusion, we have performed numerical optimal-control-theory calculations to study various aspects of quantum state preparation and unitary operator synthesis using elementary gates. These calculations allowed us to determine minimum quantum circuit sizes and depths for some few-qubit tasks. Furthermore, the flexibility afforded by numerical calculations allowed us to analyze statistical information related to all the possible gate configurations. It also allowed us to investigate the use of alternative gate sets, e.g.~ones with multi-qubit gates instead of two-qubit CNOT gates as the entangling gates in the elementary gate set.

Among the results that we found is the fact that theoretical lower bounds in the literature generally, but not always, coincide with the actual minimum numbers of gates needed for various tasks. Other interesting results include the high fidelities obtained even below the minimum number of gates for perfect task implementation and the large multiplicity of quantum circuits that lead to a perfect implementation of the target task, even at the minimum required number of gates.

Recent studies have shown that quantum circuits can be simplified and/or accelerated by the use of ancilla qubits \cite{Camps,Sun} or additional quantum states in each qubit \cite{Inada,Galda,Ashhab}. It will be interesting to extend our work to study these more complex situations. The results presented in this manuscript demonstrate that numerical methods can be a powerful tool to complement the theoretical approaches used in the literature on quantum gate decomposition. Our approach can also be applied in future studies on quantum circuit optimization, including in cases where realistic physical constraints apply to specific quantum computing devices.

We would like to thank T. Nakayama for setting up the cloud computing environment that was used for some of our calculations. This work was supported by MEXT Quantum Leap Flagship Program Grant Number JPMXS0120319794 and by Japan Science and Technology Agency Core Research for Evolutionary Science and Technology Grant Number JPMJCR1775.


\begin{thebibliography}{99}

\bibitem{Ladd} T. D. Ladd, F. Jelezko, R. Laflamme, Y. Nakamura, C. Monroe, and J. L. O'Brien, Quantum computers, Nature {\bf 464}, 45 (2010).

\bibitem{Buluta} I. Buluta, S. Ashhab, and F. Nori, Natural and artificial atoms for quantum computation, Rep. Prog. Phys. {\bf 74}, 104401 (2011).

\bibitem{Arute} F. Arute {\it et al.}, Quantum supremacy using a programmable superconducting, Nature {\bf 574}, 505 (2019).

\bibitem{Wu} Y. Wu {\it et al.}, Strong quantum computational advantage using a superconducting quantum processor, Phys. Rev. Lett. {\bf 127}, 180501 (2021).

\bibitem{Barenco} A. Barenco, C. H. Bennett, R. Cleve, D. P. DiVincenzo, N. Margolus, P. Shor, T. Sleator, J. Smolin, and H. Weinfurter, Elementary gates for quantum computation, Phys. Rev. A {\bf 52}, 3457 (1995).

\bibitem{Knill} E. Knill, Approximation by Quantum Circuits, LANL report LAUR-95-2225; arXiv:quant-ph/9508006.

\bibitem{ShendeUnitary} V. V. Shende, I. L. Markov, and S. S. Bullock, Minimal universal two-qubit controlled-NOT-based circuits, Phys. Rev. A {\bf 69}, 062321 (2004).

\bibitem{Bergholm} V. Bergholm, J. J. Vartiainen, M. M\"ott\"onen, and M. M. Salomaa, Quantum circuits with uniformly controlled one-qubit gates, Phys. Rev. A {\bf 71}, 052330 (2005).

\bibitem{Mottonen} M. M\"ott\"onen and J. J. Vartiainen, Decompositions of general quantum gates, In {\it Trends in Quantum Computing Research}, S. Shannon Ed. (NOVA Publishers, New York, 2006).

\bibitem{Nielsen} M. A. Nielsen and I. L. Chuang, {\it Quantum Computation and Quantum Information} (Cambridge University Press, New York, 2000).

\bibitem{Grover} L. K. Grover, Synthesis of quantum superpositions by quantum computation, Phys. Rev. Lett. {\bf 85}, 1334 (2000).

\bibitem{Kraus} B. Kraus and J. I. Cirac, Optimal creation of entanglement using a two-qubit gate, Phys. Rev. A 63, 062309
(2001).

\bibitem{Kaye} P. Kaye and M. Mosca, Quantum networks for generating arbitrary quantum states, Proceedings of the International Conference on Quantum Information (Rochester, New York, 2001).

\bibitem{ShendeStatePrep} V. V. Shende and I. L. Markov, Quantum Circuits For Incompletely Specified Two-Qubit Operators, Quant. Inf. and Comp. {\bf 5}, 49 (2005).

\bibitem{Soklakov} A. N. Soklakov and R. Schack, Efficient state preparation for a register of quantum bits, Phys. Rev. A {\bf 73}, 012307 (2006).

\bibitem{Znidaric} M. Žnidarič, O. Giraud, and B. Georgeot, Optimal number of controlled-NOT gates to generate a three-qubit state, Phys. Rev. A {\bf 77}, 032320 (2008).

\bibitem{Plesch} M. Plesch and C. Brukner, Quantum-state preparation with universal gate decompositions, Phys. Rev. A {\bf 83}, 032302 (2011).

\bibitem{Vidal} G. Vidal and C. M. Dawson, Universal quantum circuit for two-qubit transformations with three controlled-not gates, Phys. Rev. A {\bf 69}, 010301(R) (2004).

\bibitem{Vatan} F. Vatan and C. Williams, Optimal quantum circuits for general two-qubit gates, Phys. Rev. A {\bf 69}, 032315 (2004).

\bibitem{Vartiainen} J. Vartiainen M. M\"ott\"onen, and M. M. Salomaa, Efficient decomposition of quantum gates, Phys. Rev. Lett. {\bf 92}, 177902 (2004).

\bibitem{GoubaultDeBrugiere2020} T. Goubault de Brugière, M. Baboulin, B. Valiron, and C. Allouche, Quantum circuit synthesis using Householder transformations, Computer Physics Communication {\bf 248}, 107001 (2020).

\bibitem{Rakyta} P. Rakyta and Z. Zimbor\'as, Approaching the theoretical limit in quantum gate decomposition, Quantum {\bf 6}, 710 (2022).

\bibitem{GoubaultDeBrugiere2019} T. Goubault de Brugière, M. Baboulin, B. Valiron, and C. Allouche, Synthesizing quantum circuits via numerical optimization, Proceedings of ICCS, LNCS 11537, 3, (2019).

\bibitem{Martinez} E. A. Martinez, T. Monz, D. Nigg, P. Schindler, and R. Blatt, New J. Phys. {\bf 18}, 063029 (2016).

\bibitem{Cerezo} M. Cerezo, K. Sharma, A. Arrasmith, and P. J. Coles, Variational quantum state eigensolver, arXiv:2004.01372.

\bibitem{Shirakawa} T. Shirakawa, H. Ueda, and S. Yunoki, Automatic quantum circuit encoding of a given arbitrary quantum state, arXiv:2112.14524.

\bibitem{Stewart} G. W. Stewart, The efficient generation of random orthogonal matrices with an application to condition estimators, SIAM Journal on Numerical Analysis {\bf 17}, 403 (1980).

\bibitem{Khaneja} N. Khaneja, T. Reiss, C. Kehlet, T. S. Herbr\"uggen, S. J. Glaser, Optimal control of coupled spin dynamics: design of NMR pulse sequences by gradient ascent algorithms, J. Magn. Reson. {\bf 172}, 296 (2005).

\bibitem{ConvergenceFootnote} The reliability in generating high-fidelity results suggests that there are no local minima in the landscape of the fidelity as a function of control parameters. In this situation, increasing the number of iterations should always allow the algorithm to approach the maximum achievable fidelity. We note, however, that our calculations exhibited variations in the speed of convergence. For example, when we took $N=6$ and $10^4$ iterations, the majority of the $F=1$ configuration permutations gave $1-F<10^{-8}$, but a few of them gave $1-F\sim 10^{-5}$-$10^{-2}$. We found that trying different (random) initial guesses for the single-qubit rotations is sometimes a more efficient (i.e.~faster) way to reach the maximum value of $F$ than increasing the number of iterations. We suspect that the reason for this situation is a technical one related to the way that we set up our calculations: we decrease the step size when the fidelity seems to be approaching its maximum value, which occasionally makes the convergence extremely slow.

\bibitem{Camps} D. Camps and R. Van Beeumen, Approximate quantum circuit synthesis using block encodings, Phys. Rev. A {\bf 102}, 052411 (2020).

\bibitem{Saeedi} M. Saeedi and M. Pedram, Linear-depth quantum circuits for n-qubit Toffoli gates with no ancilla, Phys. Rev. A {\bf 87}, 062318 (2013).

\bibitem{Sun} X. Sun, G. Tian, S. Yang, P. Yuan, and S. Zhang, Asymptotically optimal circuit depth for quantum state preparation and general unitary synthesis, arXiv:2108.06150.

\bibitem{Inada} T. Inada, W. Jang, Y. Iiyama, K. Terashi, R. Sawada, J. Tanaka, and S. Asai, Measurement-free ultrafast quantum error correction by using multi-controlled gates in higher-dimensional state space, arXiv:2109.00086.

\bibitem{Galda} A. Galda, M. Cubeddu, N. Kanazawa, P. Narang, and N. Earnest-Noble, Implementing a ternary decomposition of the Toffoli gate on fixed-frequency transmon qutrits, arXiv:2109.00558.

\bibitem{Ashhab} S. Ashhab F. Yoshihara, T. Fuse, N. Yamamoto, A. Lupascu, and K. Semba, Speed limits for two-qubit gates with weakly anharmonic qubits, Phys. Rev. A {\bf 105}, 042614 (2022).

\end{thebibliography}
\end{document}